\documentstyle[eqsecnum,aps,multicol,epsfig]{revtex}

\begin{document}
\draft
%\preprint{BORIS}
\title{Giant change in IR light transmission in 
La$_{0.67}$Ca$_{0.33}$MnO$_{3}$ film near the Curie temperature: promising 
application in optical devices
}
\author{Yu. P. Sukhorukov,$^{a}$, E. A. Gan'shina$^{b}$, 
B. I. Belevtsev$^{c}$,  N. N. Loshkareva$^{a}$,\\
 A. N. Vinogradov$^{b}$, K. D. D. Rathnayaka$^{d}$, A. Parasiris$^{d}$,
D. G. Naugle$^{d}$}
\address{$^{a}$Institute of Metal Physics, Ural Division of the Russian 
Academy of Sciences, 620219 Ekaterinburg, Russia\\
$^{b}$Physics Department, Moscow State University, 119899, Moscow, Russia\\
$^{c}$B. Verkin Institute for Low Temperature 
Physics \& Engineering, Ukrainian Academy of Sciences, Kharkov, 61103, 
Ukraine\\
$^{d}$Department of Physics, Texas A\&M University, 
College Station, TX 77843-4242, USA
}
%\date{\today}
\maketitle
\begin{abstract}
Transport, magnetic,  magneto-optical (Kerr effect) and optical (light 
absorption) properties have been studied in an oriented polycrystalline 
La$_{0.67}$Ca$_{0.33}$MnO$_{3}$ film which shows colossal magneto-resistance. 
The correlations between these properties are presented. 
A giant change in IR light transmission (more 
than a 1000-fold decrease) is observed on crossing the Curie temperature 
(about 270 K) from high to low temperature. Large changes
in transmittance in a magnetic field were observed as well. The giant 
changes in transmittance and the large magneto-transmittance can
be used for development of IR optoelectronic devices 
controlled by thermal and magnetic fields. Required material 
characteristics of doped manganites for these devices are discussed.
\end{abstract}
%\narrowtext
\begin{multicols}{2}
\section{Introduction}
It is known\cite{coey,dagotto} that the conductivity of doped manganites 
increases dramatically at the transition from the paramagnetic to the
ferromagnetic 
state. In the La$_{1-x}$Ca$_{x}$MnO$_3$ system with doping levels in 
the range $0.17 < x < 0.5$  this transition occurs simultaneously with an
insulator-metal transition. If the manganite samples have fairly good crystalline 
order, huge temperature changes in light absorption and 
magneto-absorption can occur at the transition to the metallic ferromagnetic 
state. These effects open the possibility for use of doped manganites not 
only as magnetic recording media, but also for developing 
various optoelectronics devices controlled by magnetic or thermal 
fields\cite{loshka1}. Previously large temperature changes in light 
absorption and magneto-absorption near the Curie temperature, $T_c$, have been 
found in single-crystal (La$_{1-x}$Pr$_x$)$_{0.7}$Ca$_{0.3}$MnO$_3$ films 
($T_c \approx 185$~K) \cite{loshka1} and single crystal 
La$_{0.9}$Sr$_{0.1}$MnO$_3$ ($T_c \approx 160$~K) \cite{loshka2}. This work
reports similar effects in polycrystalline
La$_{0.67}$Ca$_{0.33}$ MnO$_3$ films with a higher $T_c$, which is only
moderately
less than room temperature.  This may be important in practical 
use of these effects. This study will also address correlations of other
material properties (the structural, transport 
and magnetic properties) with the optical measurements for this system. 

\section{Experiment}
The La$_{1-x}$Ca$_{x}$ MnO$_3$ ($x \approx 1/3$) film described in this paper 
was grown by pulsed-laser deposition (PLD) on a (001) oriented 
LaAlO$_3$ substrate. A PLD system from 
Neocera Inc. with a Lambda Physik KrF excimer laser operating at 248~nm was 
used to ablate the target material with a nominal composition 
La$_{2/3}$Ca$_{1/3}$MnO$_{3}$. The target was prepared by solid-state 
reaction starting from high purity powders of La$_2$O$_3$, CaCO$_3$ and
MnCO$_3$. 
The lattice parameter for the target indexed for a pseudo-cubic unit cell is 
found to be $a_p = 0.38435$~nm. The film (about 150 nm thick) was ablated at 
a substrate temperature of 600$^{\circ}$C in an oxygen atmosphere at 
pressure $P_{O2} = 250$~mTorr. During deposition the pulse energy  was 
584 mJ with a repetition rate of 8 Hz. The target-substrate distance was 
about 7 cm. Time of deposition was about 31~min. After deposition the film
was cooled to room temperature in the same oxygen atmosphere. The film was 
post-annealed in flowing oxygen for 25 hours at 950$^{\circ}$C.  
\par
The film characterization and measurements were done using a variety of
experimental techniques. A standard $\Theta-2\Theta$ scan was used for
the X-ray diffraction (XRD) study of the film. The XRD patterns were obtained
using a Rigaku model D-MAX-B diffractometer with a graphite monochromator 
and Cu-K$_{\alpha1,2}$ radiation. The AC susceptibility and DC magnetization
were measured in a Lake Shore model 7229 AC Susceptometer/DC 
Magnetometer. Resistance as a function of temperature and magnetic field was
measured using a standard four-point probe technique in magnetic fields up to
5 T.
\par
Optical absorption spectra and temperature dependences of the intensity of 
the transmitted light (transmission) were investigated in the energy range 
0.12--1.0 eV, the temperature range 80-295 K and in magnetic fields up 
to 0.8 T. The constant
magnetic field was applied along the direction of light propagation 
perpendicular to the plane of the film. The magneto-optical (MO) properties 
of the film were studied by measurements of the linear transverse Kerr effect
(TKE). The TKE was investigated in the energy range 0.5--3.8 eV, the
temperature range 10--300 K and in magnetic fields 0.3 T.  A dynamic 
method to record TKE was used. The relative change in the intensity of the
reflected light $\delta = [I(H)-I(0)]/I(0)$, where $I(H)$ and $I(0)$ are
the intensities of the reflected light in the presence and in absence of 
a magnetic field respectfully, was directly measured in the experiment.
Other
details of the technique used were described previously\cite{balukina}. 

\section{Results and discussion}

\subsection{Structural, transport and magnetic properties}

A $\Theta-2\Theta$ scan permits determination of the lattice parameters 
in the direction perpendicular to the film plane. In the XRD 
pattern (Fig. 1) only the (001), (002), (003) and (004) reflections from both
the substrate and the film are found, indicating a high degree
out-of-plane orientation of the film. The XRD pattern for the substrate 
corresponds to single-crystal LaAlO$_3$.  At fairly high angles the 
peaks for the substrate are split (having doublets), due to contribution 
of the K$_{\alpha 1}$ and K$_{\alpha 2}$ radiation. The less intense film peaks 
are not split. The lattice parameter for the substrate is found to be 
$a_s = 0.3789 \pm 0.0002$~nm. This agrees well with the expected value 
$a = 0.37896$~nm for single-crystal LaAlO$_3$\cite{bueble}. The
out-of-plane lattice parameter for the film is equal to 
$a_f = 0.3838\pm 0.0002$~nm which is a little less than that of the target. 
The surface of the film appears bright and specular.  Other films having the
same structural, magnetic and transport  properties were prepared in the same
PLD run.  The rms roughness of one of these films was determined to be less than
2 nm in an atomic force microscope.  The film studied here is expected to have
has the same high surface smoothness.
\par
The temperature dependence of the real ($\chi^{'}$) and the imaginary 
($\chi^{''}$) components of the AC susceptibility are shown in Fig.~2. 
In the same figure the temperature dependence of the magnetization is presented.
The latter is rather noisy due to the small mass of the film.  Nonetheless,
it is evident from Fig. 2 that the paramagnetic-ferromagnetic 
transition is quite sharp. From these temperature dependences the Curie 
temperature, $T_c$, can be obtained.  The value of 
$T_c\approx 276$~K, is found if $T_c$ is defined as the temperature of the
inflection point in the $\chi^{'} (T)$ curve. It is expected also that
$\chi^{''}(T)$ should peak at $T_c$\cite{ara}. This temperature is
about 273.3~K, which is very close to the $T_c$ value found from the 
$\chi^{'} (T)$ curve. 
\par
The temperature dependence of the film resistivity and of the magneto-resistance 
(MR) are shown in Figs. 3 and 4.  As evident from Fig. 3, the resistive 
transition which accompanies the magnetic one is also sharp (see
also the temperature behavior of the temperature coefficient of resistance
(TCR) shown in the insert to this Figure). In general, the observed 
$\rho (T)$ behavior 
(as well as the $\chi^{'} (T)$ curve) reflects the high-quality 
of the film studied. The temperature, $T_p = 277.5$~K, 
where resistance peaks, is very close to $T_c$, as expected for high quality 
films\cite{ara}. The ratio of resistances at $T_p$ and 4.2 K, $R_p/R_{4.2}$, 
is extremely high (about 48.6). This large variation of resistance 
with temperature is attributed mainly to the strengthening of the magnetic 
order with decreasing temperature.  The resistivity at $T=4.2$~K is about 
0.9 $\mu \Omega$~m which is one of the lowest values reported for 
La$_{1-x}$Ca$_{x}$ MnO$_3$ ($x \approx 1/3$) films (e.g., compared with those in
Ref. \onlinecite{rho}).  As a measure of the MR 
$\delta_{H} = [R(H)-R(0)]/R(0)$ is plotted in Fig. 4.  The MR has a sharp 
maximum (with an absolute value of 0.65 at $H=5$~T) at $T_m=275.6$~K, that 
is actually at $T= T_c$. Nearly identical and high values of $T_c$, $T_p$ 
and $T_m$ found in this study provide evidence of the high quality
of this highly oriented film.

\subsection{Magneto-optical and optical properties} 
The spectral dependence of TKE for the film studied (Fig. 5a) is consistent 
with previous data for doped manganites with similar doping level 
\cite{eagan,eagan2}. There is a large ``negative peak'' at $E \approx 2.8$~eV 
together with an additional feature near $E\approx 1.6$~eV. At 
$T=55$~K, the peak amplitude near $E= 2.8$ eV is about $-23\times 10^{-3}$. 
The temperature dependence of TKE (Fig.~5b) was measured at two
incident photon energies (1.8~eV and 2.8 eV), corresponding to the peak
positions
in TKE spectra (Fig. 5a).  These temperature dependences should reflect that 
of the magnetization. Indeed, the temperature at which an appreciable
Kerr effect can be seen, when going from high to lower temperature, is close 
to the $T_c$ values obtained from the magnetic and resistance measurements. On
the other 
hand, the temperature dependence curves for TKE recorded at different photon
energies, are quite different from each other.  It can be seen that 
temperature dependence of TKE, measured at $E=1.8$~eV, looks quite similar 
to the M(T) dependence 
(Fig.~2b).  Both curves reveal a sharp increase in magnetization near $T_c$, but
that for $E=2.8$~eV, shows
a far slower increase in the TKE signal with decreasing temperature below
$T_c$. This effect is most likely associated with the shifting of the peak to
lower energy (``red'' region) with increasing temperature 
(Fig. 5a); whereas the peak at $E=1.8$~eV is not shifted for a
wide temperature range.  The mechanism of this difference in temperature 
behavior of the spectral peaks is not clear at present. 
It is conceivable that this is connected with the different nature of optical 
transitions responsible for magneto-optical activity in these energy regions.
As was shown earlier \cite{eagan3}, in the spectral range under study there 
is an allowed electric-dipole transition in the octahedral complex 
(MnO$_{6})^{9-}$ at $E = 3.5$~eV and spin-resolved $d-d$ transition  in
Mn$^{3+}$ and Mn$^{4+}$ at lower energies. That is, a large 
magneto-optical effect in the neighborhood of $E=3$~eV is caused by a
charge-transfer transition with an involvement of both  Mn and O ions.
The temperature dependence of this transition should be more complex
than that for Mn ions, as is actually observed in this study.
Further studies are necessary, of course, to clarify this effect. In 
general, the TKE data obtained correlate well with the results of magnetic 
and transport measurements and support arguments that this film is of excellent
quality and homogeneity.
\par
Let us turn now to the optical properties of the film. An absorption 
spectrum for the film in the paramagnetic state at $T=295$~K is presented in 
Fig. 6. It can be seen that an increase in wavelength, $\lambda$, 
leads to a decrease in absorption, that can be associated with the influence 
of ``tails'' of inter-band optical transitions. For a large enough increase in 
wavelength, the absorption begins to increase. This is determined by the 
presence of the impurity band, which is usually positioned around 
$\lambda = 10$~$\mu$m ($E=0.12$~eV) for manganites and associated with Mn$^{4+}$
ions \cite{loshka1,loshka2,loshka3}.  Below $T_c$ a strong increase in 
absorption has 
been observed. This is connected with the light absorption by free charge 
carriers.  This phenomena is demonstrated more clearly by the temperature 
dependence of intensity of light transmitted through the film 
at a wavelength $\lambda =6.4$~$\mu$m (Fig. 7).
\par
The light intensity is sharply reduced below 270 K in a narrow temperature 
range (Fig.~7). The giant change in the intensity is more than a 1000-fold 
decrease. The magnetic field shifts the intensity curve $I(T)$ to higher 
temperature. The relative change of transmission under the influence of the
magnetic
field [magneto-transmittance (MT)],
$\Delta I_{H}/I_{0} = [I_{H} - I_{0}]/I_{0}$, peaks at 28 \% in a
field of 0.8 T at $T=265$~K (see insert in Fig. 7). The magneto-transmittance
is found to be significant for the film studied in the temperature range 
250-280 K.  That can be important for possible applications. In previous works 
\cite {loshka1,loshka2} this effect was observed at significantly lower
temperatures 
($T < 200$~K). It must be emphasized as well that the effect is seen in 
unpolarized light.

\subsection{Discussion}
It is quite clear that the giant decrease in transmission found below $T_c$ in 
this study is determined by the transition of the film to the 
metallic state with fairly high density of free 
(or quasi-free) charge carriers. In spite of extensive experimental and
theoretical efforts, a clear understanding of the metal-insulator transition in
doped manganites and the colossal magneto-resistance (CMR) associated with it
is not yet available.  Several possible explanations can be found in review 
papers \cite{coey,dagotto}.  One of the possible reasons for CMR in 
homogeneous samples is the shift of the mobility edge induced by a change 
in temperature or magnetic field \cite{bebenin}.  
However, it is hard to expect a sufficiently high homogeneity in doped 
manganites \cite{dagotto}.  One of the certain things is 
that a clear correlation exists between transport properties and 
magnetization in the mixed-valence manganites \cite{coey,dagotto}.
Namely, the resistance $R$ of manganites in the ferromagnetic state is a 
function of magnetization and the conductivity increases with enhancement 
of ferromagnetic order. This, together with the large drop of resistivity and
light transmission below $T_c$, is the source of the huge negative MR and MT
in manganites. 
\par
The sharp drop in resistance below $T_c$ in doped manganites can be used for
bolometric applications \cite{goyal}. Actually, the TCR of the film studied
is fairly high in the temperature range close to room temperature (see insert
in Fig. 3). It has been shown here, however, that changes in the transmission at
the magnetic transition are much larger. Indeed, the 
ratio of resistances at $T_p$ and 4.2 K, $R_p/R_{4.2}$, was found to be 
very high (about 48.6), but the optical transmission exhibits a far 
larger drop (more than a 1000-fold decrease) below $T_c$.
\par
The transmission, $t$, of films is defined as the ratio between intensities  
of transmitted and incident light. For a film with high absorption , when 
$t < 10$~\%, the transmission is related to the absorption coefficient, $K$, 
by the equation $t=(1-R_{1})(1-R_{2})(1-R_{12})\exp (-Kd)$, where $d$
is film thickness, $R_{12}$, $R_{2}$, $R_{1}$ are reflection coefficients at the
boundaries of film-substrate, substrate-vacuum, film-vacuum, respectfully.
The absorption coefficient is proportional to the DC 
conductivity (Drude model). It is evident from the foregoing equation that
there is no direct proportionality between the transmission and the
resistivity, but the temperature dependence of the transmission reflects the 
$\rho (T)$ behavior. 
The change in intensity of the transmitted light through the sample at the
transition to the ferromagnetic state is of particular interest for practical
use. 
This change is large for the film studied in comparison with the moderate 
change in the absorption coefficient. (It can be shown that a 1000-fold decrease
in the transmission corresponds to an increase in the absorption 
coefficient by a factor of 5.6, not taking into account the change in 
reflectivity.) The decrease in skin depth, $\delta$, at $T<200$~K is 
responsible for the non-transparency of the film in the ferromagnetic state.  
This can be estimated with the expression 
$\delta = (2\rho /\mu \mu_{0} \omega )^{1/2}$ under the assumption that the
magnetic
permeability is equal to unity to obtain $\delta \approx 70$~nm at $T=4.2$~K,
and $\delta \approx 486$~nm at $T=T_p$ for light with   
$\lambda = 6.4$~$\mu$m.  That is, the skin depth is far larger 
than the film thickness ($d \approx 150$~nm) in  the paramagnetic state,
but in the ferromagnetic state it is significantly less than the film thickness.
It should be added  that the relative change of the intensity of 
transmitted light in a magnetic field, MT = $[I_{H} - I_{0}]/I_{0}$, is certain
to be a direct analogue of MR = ($[R(H)-R(0)]/R(0)$).  As in the case of MR, 
this quantity MT is determined  by the ability of an external magnetic field 
to enhance the magnetic order in the doped manganites.  
\par
The high-temperature optical phenomena of magneto-transmittance and the
strong temperature dependence of magneto-transmittance near $T_c$ that has 
been reported here can be used for development of a number of IR 
optoelectronic devices, such as light modulators, optical attenuators,
light shutters, temperature and magnetic-field indicators, etc.  These devices
can work in the IR range with $\lambda$ up to 14 
$\mu$m. As far as we know, the number of materials, which can be used in 
this energy range as optical control devices, is very limited.
High quality and high homogeneity doped manganite films which 
exhibit a sharp magnetic transition and high conductivity in the ferromagnetic 
state should be very useful in development of appropriate optical 
devices.
\par
In conclusion, at a temperature ($T\approx 270$~K) not far from 
room temperature giant temperature changes in light
transmittance and large magneto-transmittance has been observed. These effects
can be used for IR optoelectronics devices.
\acknowledgments
This study was supported in part by INTAS (project INTAS-97-30253) and by the
Ministry of Science of Russia (contract No. 2.4.01). The research at 
Texas A\&M University was supported by the Robert A Welch Foundation 
(A-0514) and the Texas Center for Superconductivity of the University of
Houston (TCSUH).

\begin{figure}
\centerline{\epsfig{file=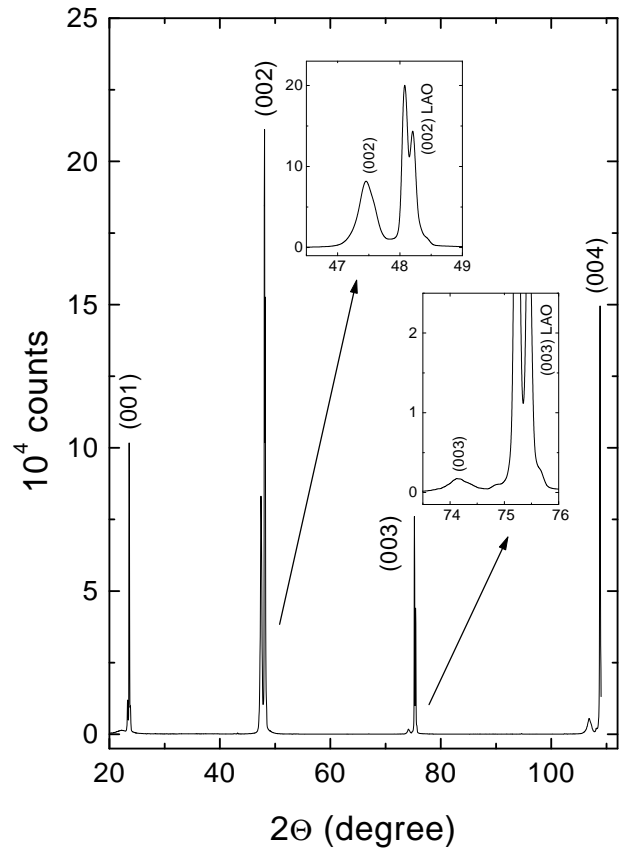,width=9.0cm}}
\caption{XRD $\Theta -2\Theta$ curve of the La$_{0.67}$Ca$_{0.33}$MnO$_{3}$ 
film on a LaAlO$_{3}$ (LAO) substrate.  The indexes (001), (002), 
(003) and (004) (of a pseudo-cubic unit cell)  apply to the regions where
reflections with these indexes are located for both for the film and the 
substrate. The regions around the (002) and (003) reflections 
are magnified in inserts.} 
\label{Fig.1}
\end{figure}

\begin{figure}
\centerline{\epsfig{file=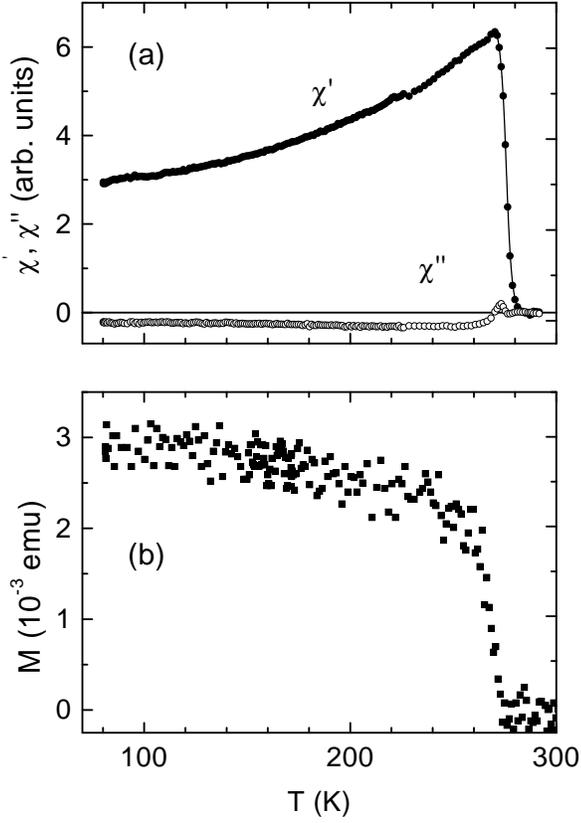,width=9.0cm}}
\caption{Temperature dependences of the AC susceptibility (a) and the
magnetization (b).  The real ($\chi^{'}$) and imaginary
($\chi^{''}$) parts of AC susceptibility were  recorded in AC magnetic 
field $H_{AC} = 0.2$~mT at frequency 125 Hz for field direction parallel to 
the film plane. $M(T)$ dependence was obtained at $H_{DC} = 2.5$~mT. 
All dependences were recorded with temperature increasing
after the film was cooled in zero field.}
\label{Fig.2}
\end{figure}

\begin{figure}
\centerline{\epsfig{file=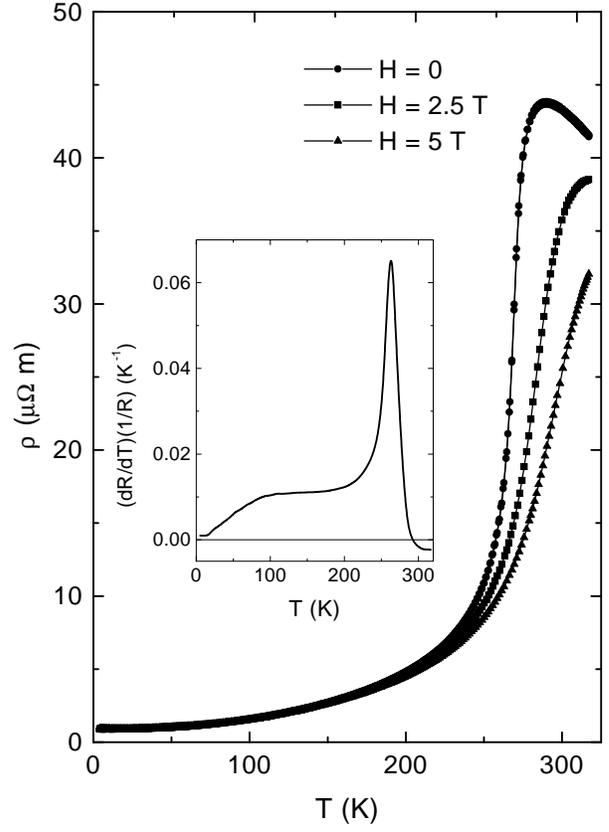,width=9.0cm}}
\caption{Temperature dependence of the resistivity of the investigated
film in zero magnetic field and in fields $H=2.5$~T and $H=5$~T. Field $H$
was perpendicular to the film plane. The insert shows the temperature
dependence of the temperature coefficient of resistance (TCR) at $H=0$.}
\label{Fig.3}
\end{figure}

\begin{figure}
\centerline{\epsfig{file=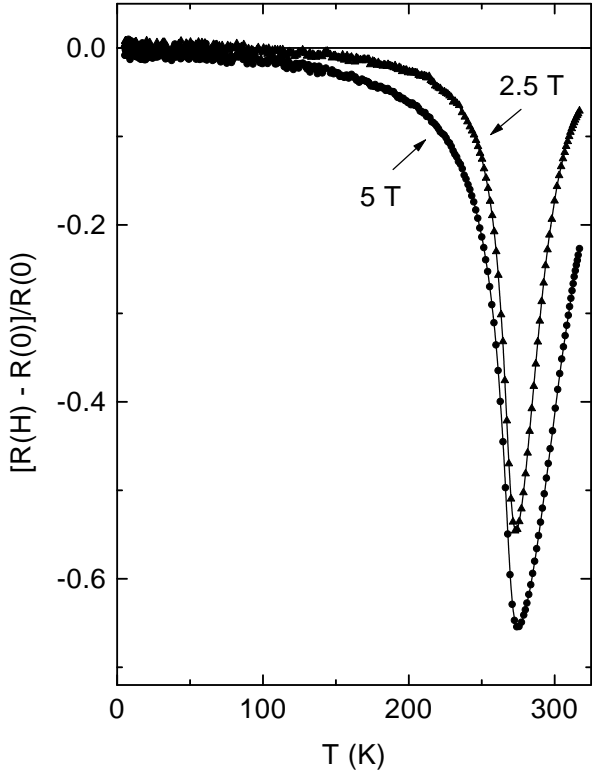,width=9.0cm}}
\caption{Temperature dependence of the magneto-resistance $[R(H)-R(0]/R(0)$ 
at $H=2.5$~T and $H=5$~T. Field $H$ was perpendicular to the
film plane.}
\label{Fig.4}
\end{figure}

\begin{figure}
\centerline{\epsfig{file=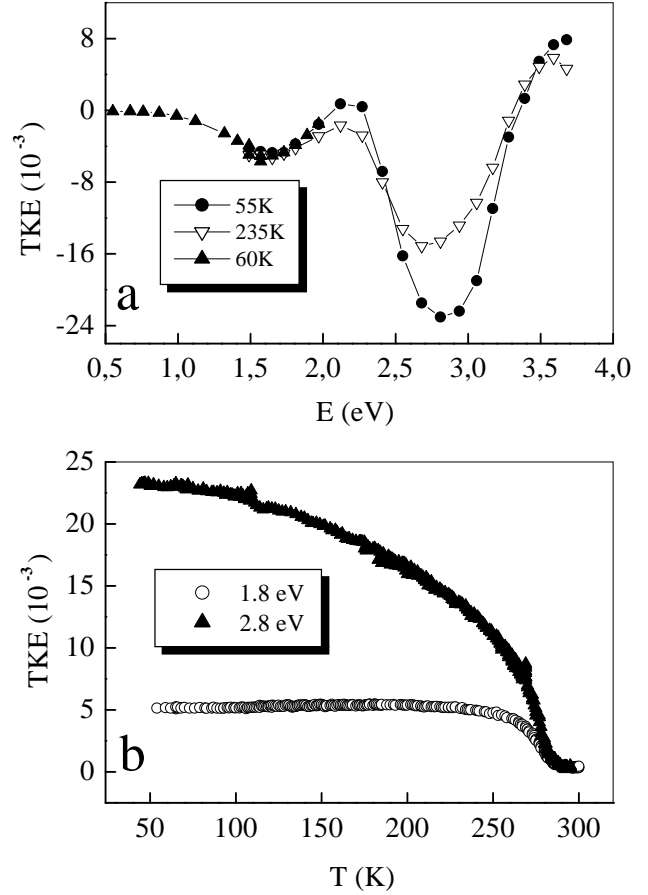,width=9.0cm}}
\caption{Spectral dependences of TKE at different temperatures (a) 
and temperature dependences of TKE at different photon energies (b).
All dependences were recorded at light incidence angle $\phi = 68^{\circ}$.}
\label{Fig.5}
\end{figure}

\begin{figure}
\centerline{\epsfig{file=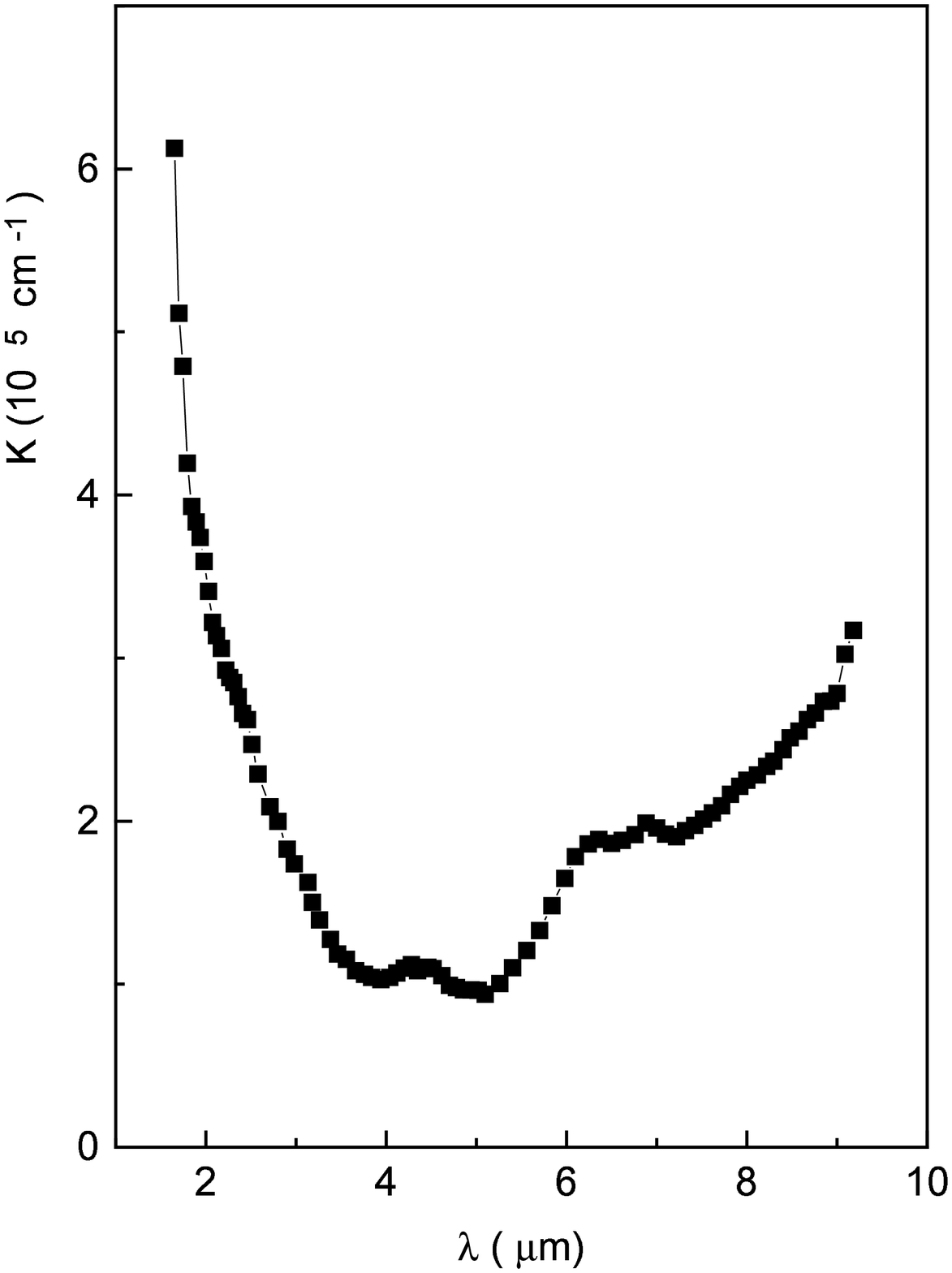,width=8.6cm}}
\caption{The optical absorption spectrum of 
La$_{0.67}$Ca$_{0.33}$MnO$_{3}$ film at $T=295$~K.}
\label{Fig.6}
\end{figure}

\begin{figure}
\centerline{\epsfig{file=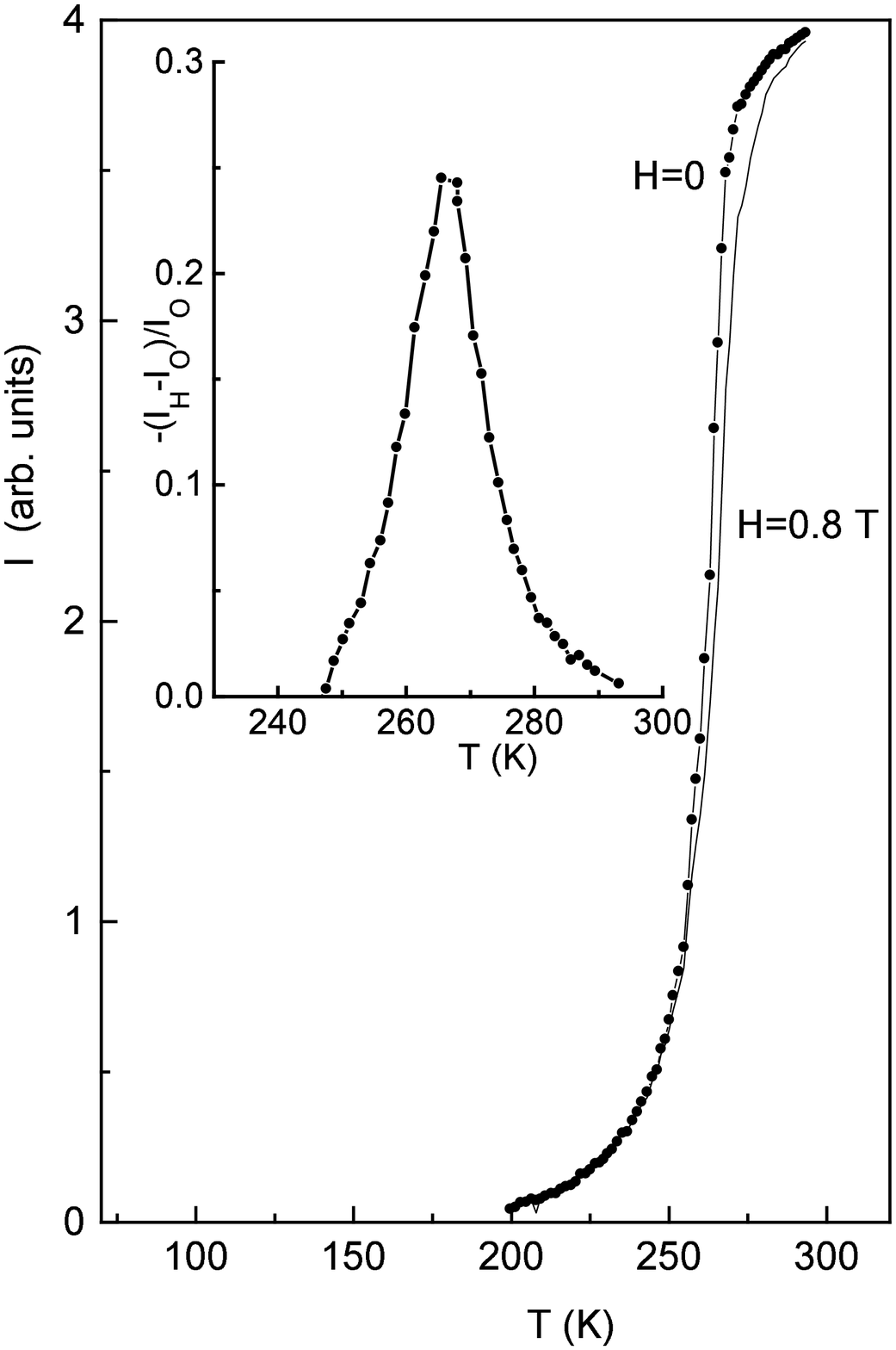,width=8.8cm}}
\vspace{22pt}
\caption{Temperature dependences of the intensity of light transmitted
through La$_{0.67}$Ca$_{0.33}$MnO$_{3}$ film in zero magnetic field and at 
field $H=0.8$~T for wavelength $\lambda = 6.4$~$\mu$m. The insert 
shows the temperature dependence of magneto-transmittance at $H=0.8$~T for 
the same wavelength.}
\label{Fig.7}
\end{figure}

\end{multicols}

\begin{references}

\bibitem{coey}J. M. D. Coey, M. Viret, and S. von Molnar, Adv. Phys.
{\bf 48,} 167 (1999).

\bibitem{dagotto}E. Dagotto, T. Hotta, A. Moreo, Phys. Rep. {\bf 344,}
1 (2001).

\bibitem{loshka1}Yu. P. Sukhorukov, N. N. Loshkareva, E. A. Gan'shina,
A. R. Kaul, O. Yu. Gorbenko, K. A. Fatieva, Techn. Phys. Lett. 
{\bf 25,} 551 (1999).

\bibitem{loshka2}N. N. Loshkareva, Yu. P. Sukhorukov,  B. A. Gizhevskii,
A. A. Samokhvalov, V. E. Arkhipov, V. E. Naish, S. G. Karabashev, 
Ya. M. Mukovskii, phys. stat. solidi A {\bf 164,} 863 (1997).

\bibitem{balukina}E. A. Balukina, E. A. Gan'shina, and G. S. Krinchik,
Zh. Eksp. Teor. Fiz. {\bf 93,} 1879 (1987).

\bibitem{bueble}S. Bueble, K. Knorr, E. Brecht, W. W. Schmahl, Surf. Sci.
{\bf 400,} 345 (1998).

\bibitem{ara} F. M. Araujo-Moreira, M. Rajeswari, A. Goyal, K. Ghosh,
V. Smolyaninova, T. Venkatesan, C. J. Lobb, and R.L. Greene, Appl. Phys. 
Lett. {\bf 73,} 3456 (1998).

\bibitem{rho}C. Osth\"over, K. Shmidt, and R. R. Arons, Mat. Sci. Eng. B
{\bf 56,} 164 (1998); F. Martin, G. Jakob, W. Westerburg, H. Adrian, J. 
Magn. Magn. Mat. {\bf 196-197,} 509 (1999);
R. B. Praus, B. Leibold, G. M. Gross, H.-U. Habermeier, Appl. Surf. Sci 
{\bf 138-139,} 40 (1999).

\bibitem{eagan}E. A. Gan'shina, O. Yu. Gorbenko, A. G. Smechova, 
A. R. Kaul, N. A. Babushkina, and L. M. Belova, J. Phys.: Condens. Matter
{\bf 12,} 2857 (2000).

\bibitem{eagan2}E. A. Gan'shina, O. Yu. Gorbenko, A. G. Smechova, 
A. R. Kaul, N. A. Babushkina, and L. M. Belova, in {\it Non-Linear 
Electromagnetic Systems}, edited by V. Kose and J. Sievert 
(IOS Press, Amsterdam, 1998), p.325. 

\bibitem{eagan3}E. A. Balukina, E. A. Gan'shina, G. S. Krinchik, and
A. Yu. Trifonov, J. Magn. Magn. Mater. {\bf 117,} 259 (1992).

\bibitem{loshka3}N. N. Loshkareva, Yu. P. Sukhorukov,  E. A. Neifel'd,
V. E. Arkhipov, A. V. Korolev, V. S. Gaviko, E. V. Panfilova, V. P. Dyakina,
Ya. M. Mukovskii, and D. A. Shulyatev, JETP {\bf 90,} 389 (2000).

\bibitem{bebenin}N. G. Bebenin and V. V. Ustinov, J. Magn, Magn. Mater. 
{\bf 196-197,} 451 (1999).

\bibitem{goyal}A. Goyal, M. Rajeswari, R. Shreekala, S. E. Lofland, 
S. M. Bhagat, T. Boettcher, C. Kwon, R. Ramesh, and T. Venkatesan,
Appl. Phys. Lett. {\bf 71,} 2535 (1997).

\end{references}
\end{document}